\documentclass[12pt]{article}
\begin{document}
\title{THE LORENTZ-DIRAC AND DIRAC EQUATIONS}
\author{B.G. Sidharth\\
International Institute for Applicable Mathematics \& Information Sciences\\
Hyderabad (India) \& Udine (Italy)\\
B.M. Birla Science Centre, Adarsh Nagar, Hyderabad - 500 063 (India)}
\date{}
\maketitle
\begin{abstract}
It is well known that the Classical theory of the electron reached the limits of its description at time intervals of the order of $10^{-23} secs$, that is the Compton time. It is widely believed that below these time intervals Classical Electrodynamics doesn't work and that a Quantum description is required. Using the Lorentz Dirac and the Dirac equations of the electron, we point out that in fact there is a convergence of the two descriptions at the Compton scale.
\end{abstract}
\section{Introduction}
In the thirties when Dirac was working on the relativistic Quantum Theory of the electron, he was also working on the theory in the Classical description. Indeed the limitation of the classical description of the electron appears in the infinity we obtain when the size of the electron $\to 0$ as also in the radiation reaction term (also called the Schott term) which contains the third derivative \cite{rohr,barut,feyn}. Herein was the first ingredient of what later became renormalization in electrodynamics. In fact we have the Lorentz Dirac equation,
\begin{equation}
ma^\mu = F^\mu_{in} + F^\mu_{ext} + \Gamma^\mu\label{6-57}
\end{equation}
with
\begin{equation}
F^\mu_{in} = \frac{e}{c} F^{\mu \nu}_{in} v_\nu\label{6-56}
\end{equation}
and
\begin{equation}
F^\mu = \frac{2}{3} \frac{e^2}{c^3} \left(\dot{a}^\mu - \frac{1}{c^2} a^\lambda a_\lambda v^\mu\right)\label{6-55}
\end{equation}
where $a^\mu$ denotes the acceleration of the electron and
\begin{equation}
\frac{2}{3} \frac{e^2}{c^3} \dot{a}^\mu\label{x}
\end{equation}
which appears in (\ref{6-55}) is the Schott term. It must be mentioned that $m$ contains the electromagnetic mass,
\begin{equation}
m_{elm} = \frac{e^2}{8\pi c^3} \int^\infty_R \frac{1}{r^4} r^2 dr d\Omega = \frac{e^2}{2Rc^2},\label{6-22}
\end{equation}
where, in (\ref{6-22}) $R$ is the radius of the electron treated as a sphere. It must be mentioned that (\ref{6-57}) needs to be supplemented with the condition
\begin{equation}
lim_{r\to \infty} a^\mu (\tau) = 0\label{6-37}
\end{equation}
that is that the acceleration of the electron $\to 0$ at an infinite time in the future. This overcomes the well known problem of the run away solutions of the Lorentz Dirac equation (\ref{6-57}). We will return to this briefly. It must further be noted that there is a non locality, though this is within the interval given by
\begin{equation}
\tau_0 \equiv \frac{2}{3} \frac{e^2}{mc^3}\label{6-70}
\end{equation}
With this input the Lorentz Dirac equation can be written as
\begin{equation}
m(a^\mu - \tau_0 \dot{a}^\mu ) = K^\mu\label{6-x}
\end{equation}
where $K^\nu$ is given by
\begin{equation}
K^\mu (\tau) = F^\mu_{in} + F^\mu_{ext} - \frac{1}{c^2} Rv^\mu ,\label{6-71}
\end{equation}
(\ref{6-x}) can now be integrated to give
\begin{equation}
ma^\mu (\tau) = \int^\infty_0 K^\mu (\tau + \alpha \tau_0) c^{-\alpha} d\alpha\label{6-80}
\end{equation}
Let us compare (\ref{6-80}) with the more familiar Newtonian equation
\begin{equation}
ma^\mu (\tau) = F^\mu (\tau)\label{3-12}
\end{equation}
(Cf. ref.\cite{rohr,bgsiaad} for a detailed discussion). (\ref{6-80}) differs from (\ref{3-12}) in three ways. Firstly $F^\mu$ is a function of the coordinates and the velocity, and is a second order ordinary differential equation in the coordinates, whereas (\ref{6-80}) is an integro differential equation because of the fact that the acceleration also appears in it. Secondly (\ref{3-12}) is a local equation in time, unlike (\ref{6-80}) in which the acceleration depends on the force at time $\tau$ and also at all later times. Thirdly it must be noted that (\ref{6-80}) needs to be supplemented by an additional asymptotic condition viz., (\ref{6-37}).\\
Nevertheless it must be noted that the non locality in time is associated with $\tau_0$ in (\ref{6-70}) which is of the order of the Compton time and is thus very small. Explicitly this means that for a classical system we have,
\begin{equation}
ma^\mu (\tau) = K^\mu (\tau + \xi \tau_0) \approx K^\mu (\tau)\label{6-84}
\end{equation}
which shows that the local theory is approximately correct. We also notice that if (\ref{6-84}) were exact, then the difference between it and the Lorentz Dirac equation (\ref{6-57}) would be precisely the Schott term (\ref{x}). It is thus this Schott term which is responsible for the non local time dependence (via the third time derivative). On the other hand it has been shown that \cite{bgsiaad,narhoyle} the above implies that we have to consider a small region of the order of the Compton length, around the particle, in our calculations.
\section{The Electron Self Force}
It has long been known that the root of the problem of Classical Electrodynamics, which can be seen above is the Electron Self Force. To put it simply, if the electron is considered to be a charge distribution on a spherical shell, then as long as the electron is at rest (or even moving with uniform velocity), we could argue away the force of one part of the charge distribution on another part as a cancellation due to symmetry. The situation however changes when the electron begins to accelerate - then using Special Relativity, we conclude that the electron per se is no longer in equilibrium. It was Poincare who realized that additional non electromagnetic forces were required to counteract this self force. These are the so called Poincare stresses (Cf.refs.\cite{rohr,barut,feyn}).\\
For simplicity we consider the one dimensional case, then the self force is given by
\begin{equation}
F = \frac{2}{3} \frac{e}{Re^2} x'' - \frac{2}{3} x''' + \gamma \frac{e^2 R}{c^4} x'' + 0 (R^2)\label{28-9}
\end{equation}
The important features of the self force (\ref{28-9}) are, firstly the coefficient of the triple time derivative of the coordinate is independent of the size or shape of the electron and secondly the subsequent terms are positive powers of the size and would thus $\to 0$ as the radius $R \to 0$. On the contrary, the first term is the problematic term which $\to \infty$ as the size of the electron shrinks to zero. The coefficient in the first term is precisely the electromagnetic mass. Interestingly the second term above leads precisely to the Schott term.\\
It is well known that in the thirties Dirac argued that we could still eliminate the troublesome first term, while at the same time retaining the necessary second term if we consider the electromagnetic potential to be the difference of the advanced and the retarded potentials, rather than assuming it to be the retarded potential alone. In this case he showed that (\ref{28-9}) becomes
\begin{equation}
F = - \frac{2}{3} \frac{e^2}{c^3} x''' + 0 (R)\label{28-x}
\end{equation}
The interesting thing about (\ref{28-x}) is that, the infinite term of (\ref{28-9}) is absent.
\section{The Lorentz Dirac and the Dirac Equations}
With the above input, let us consider the case when the external electromagnetic field becomes vanishingly small. Then using (\ref{28-x}) we get from (\ref{6-57}), in this one dimensional case
\begin{equation}
-\frac{2}{3} \left(e^2/c^3 \right) x''' = m x''\label{B}
\end{equation}
What is very interesting is that we get a parallel equation in the Dirac Quantum Mechanical equation of the electron \cite{dirac}. In this case we have to consider the velocity operator $c \vec{\alpha}$, where in the Dirac theory as is well known the $\alpha$'s are $4 \times 4$ matrices involving the Pauli spin matrices. However Dirac noticed that the velocity of the electron in this theory, has in this case, the eigen values $\pm c$, which of course would be impossible.\\
On the other hand, it is interesting that the same situation prevails in the Classical theory. Indeed in this case we can see in a simple way that (ref.\cite{narhoyle})
$$\dot{x} \propto exp \left(\frac{3 mc^3}{2e^2} \, t\right),$$
and within time $\tau$, the electron acquires due to the run away effect, the velocity of light.\\
Dirac then went on to obtain the equation
\begin{equation}
- \hbar^2 \ddot{\alpha}_{1} = 2 \imath \hbar \dot{\alpha}_1 H\label{A}
\end{equation}
where in (\ref{A}) $H$ is the Hamiltonian and the subscript $1$ refers to the fact that we consider, let us say the $x$ component of the velocity. Before proceeding further it can be seen, remembering that $\dot{x}$ is the velocity component, that (\ref{B}) and (\ref{A}) are identical, if we use the fact that $H = mc^2$, so that $\hbar/H$ is $\tau_0$ the Compton time, except for the factor $\imath$ in (\ref{A}).\\
Thus mathematically, the Lorentz Dirac and the Dirac equations both lead to the same equations for the velocity. Next Dirac went on to argue from (\ref{A}) that the velocity has real and imaginary parts, the imaginary part being rapidly oscillating (zitterbewegung). We encounter this situation in the limit of point space time. In real life our measurements are averaged over intervals of the order of the Compton scale. In other words space time is fuzzy \cite{uof}. Once these mathematical averages are taken over the Compton time then the rapidly oscillating terms vanish and we are left with the physical velocities and momenta.\\
Let us consider the classical case (\ref{B}). In this case we get
\begin{equation}
x'' = e^{-t/\tau_0}\label{C}
\end{equation}
where $\tau_0$ is given by (\ref{6-70}). (\ref{C}) shows that the runaway, acausal acceleration $\to 0$ for $t >$ Compton time. Within the Compton time, however there is non-locality - which is otherwise well known. The Dirac condition (\ref{6-37}) is thus recovered from the usage of the Advanced potential.\\
What is very interesting is that in the classical theory of the electron too, we have encountered minimum space time intervals at the Compton scale as seen above. We can now see that Dirac's prescription of the advanced potential of the Classical theory is indeed meaningful within the Compton scale, where as is well known there is a breakdown of causality.\\
This non physical feature within the Compton scale has been elaborated upon by Weinberg \cite{weinberg}.\\
Starting with the usual light cone of Special Relativity and the inversion of the time order of events, he goes on to add, and we quote at a little length and comment upon it, ``Although the relativity of temporal order raises no problems for classical physics, it plays a profound role in quantum theories. The uncertainty principle tells us that when we specify that a particle is at position $x_1$ at time $t_1$, we cannot also define its velocity precisely. In consequence there is a certain chance of a particle getting from $x_1$ to $x_2$ even if $x_1 - x_2$ is spacelike, that is, $| x_1 - x_2 | > |x_1^0 - x_2^0|$. To be more precise, the probability of a particle reaching $x_2$ if it starts at $x_1$ is nonnegligible as long as
$$(x_1 - x_2)^2 - (x_1^0 - x_2^0)^2 \leq \frac{\hbar^2}{m^2}$$
where $\hbar$ is Planck's constant (divided by $2\pi$) and $m$ is the particle mass. (Such space-time intervals are very small even for elementary particle masses; for instance, if $m$ is the mass of a proton then $\hbar /m = 2 \times 10^{-14}cm$ or in time units $6 \times 10^{-25}sec$. Recall that in our units $1 sec = 3 \times 10^{10}cm$.) We are thus faced again with our paradox; if one observer sees a particle emitted at $x_1$, and absorbed at $x_2$, and if $(x_1 - x_2)^2 - (x_1^0 - x_2^0)^2$ is positive (but less than or $=\hbar^2 /m^2$), then a second observer may see the particle absorbed at $x_2$ at a time $t_2$ before the time $t_1$ it is emitted at $x_1$.\\
``There is only one known way out of this paradox. The second observer must see a particle emitted at $x_2$ and absorbed at $x_1$. But in general the particle seen by the second observer will then necessarily be different from that seen by the first. For instance, if the first observer sees a proton turn into a neutron and a positive pi-meson at $x_1$ and then sees the pi-meson and some other neutron turn into a proton at $x_2$, then the second observer must see the neutron at $x_2$ turn into a proton and a particle of negative charge, which is then absorbed by a proton at $x_1$ that turns into a neutron. Since mass is a Lorentz invariant, the mass of the negative particle seen by the second observer will be equal to that of the positive pi-meson seen by the first observer. There is such a particle, called a negative pi-meson, and it does indeed have the same mass as the positive pi-meson. This reasoning leads us to the conclusion that for every type of charged particle there is an oppositely charged particle of equal mass, called its antiparticle. Note that this conclusion does not obtain in nonrelativistic quantum mechanics or in relativistic classical mechanics; it is only in relativistic quantum mechanics that antiparticles are a necessity. And it is the existence of antiparticles that leads to the characteristic feature of relativistic quantum dynamics, that given enough energy we can create arbitrary numbers of particles and their antiparticles.''\\
We note however that there is a nuance here which distinguishes Weinberg's explanation from that of Dirac. In Weinberg's analysis, one observer sees only protons at $x_1$ and $x_2$, whereas the other observer sees only neutrons at $x_1$ and $x_2$ while in between, the first observer sees a positively charged pion and the second observer a negatively charged pion. Weinberg's explanation is in the spirit of the Feynman-Stuckleberg diagrams. One particle leaves $x_1$ and then travels causally to $x_2$, where $x_1$ and $x_2$ are within the Compton wavelength of the particle. But for another observer, a particle first leaves $x_2$ and travels backward in time to $x_2$.
\section{Discussion}
We can go beyond the mathematical similarities of the Classical and Quantum theories of the electron if we recognize that there is a minimum fundamental length $l$, the Compton wavelength. Within the Compton wavelength, space time is fuzzy and is physically meaningless from the point of view of the point space time theory. The infinity problem as the electron shrinks to a point is now eliminated from the classical field. This infinity exists in Quantum theory as well, as is expressed by the fact that the velocity of the point electron equals the velocity of light - we have to average over the zitterbewegung Compton wavelength region to recover meaningful physics. Moreover once the fuzzyness within the Compton length is recognized, then it is clear that the shape of the electron becomes an irrelevant factor. It might be mentioned that Snyder had introduced the concept of this minimum fundamental length way back in 1947 \cite{snyder1,snyder2}, precisely with the motivation of eliminating the infinities.\\
It may be also mentioned in this context that the author has shown that it is possible to think of the mass $m$ purely in terms of Quantum Mechanical self interacting amplitudes within the Compton scale \cite{ijpap,cu}.\\
It is interesting to note that we get the advanced potential for example in (\ref{28-x}) by replacing ${t}$ by $-t$. Remembering that all this is now considered to take place in a very small interval of the order of the Compton length, Dirac's prescription is precisely a description of the Double Weiner process, in which case time in that interval becomes non differentiable (Cf.\cite{uof} for a detailed discussion). Briefly what happens is that the forward and backward time derivatives are unequal and in an obvious notation we have
\begin{equation}
\frac{d}{dt^+} x(t) = b_+, \quad \frac{d}{dt^-}x(t) =
b_-,\label{De6d}
\end{equation}
From (\ref{De6d}) we define two new velocities
\begin{equation}
V = \frac{b_+ + b_-}{2}, \quad U = \frac{b_+ - b_-}{2}\label{De7d}
\end{equation}
It may be pointed out that in the absence of the double \index{Wiener
process}Wiener process, $U$ given in (\ref{De7d}) vanishes while $V$ gives the
usual velocity. It is now possible to introduce a complex velocity
\begin{equation}
W = V -\imath U\label{De8d}
\end{equation}
From (\ref{De8d}) we can see that it is as if the coordinate $x$ becomes complex:
\begin{equation}
x \to x + \imath x'\label{De9d}
\end{equation}
To see this in detail, let us rewrite (\ref{De7d}) as
\begin{equation}
\frac{dX_r}{dt} = V, \quad \frac{dX_\imath}{dt} = U,\label{De10d}
\end{equation}
where we have introduced a \index{complex coordinate}complex coordinate $X$ with real and imaginary parts $X_r$ and $X_\imath$, while at the same time using derivatives with respect to time as in conventional theory.\\
We can now see from (\ref{De8d}) and (\ref{De10d}) that
\begin{equation}
W = \frac{d}{dt} (X_r - \imath X_\imath )\label{De11d}
\end{equation}
That is, in this development either we use forward and backward time derivatives and the usual space coordinate as in (\ref{De6d}), or we use the derivative with respect to the usual time coordinate but introduce complex space coordinates as in (\ref{De9d}).\\
At this stage we note that a generalization of three space dimensions leads to a quarternionic description. That is we will get four coordinates instead of three and these are represented by the three Pauli spin matrices and the unit matrix. The coordinates now become non commutative, and in fact we have equations like
\begin{equation}
[x,y] = 0(l^2)\label{5-64}
\end{equation}
Equations like (\ref{5-64}) also follow from the work of Snyder already cited. On the other hand starting from (\ref{5-64}) we can not only recover the Clifford algebra of the Dirac matrices, but also the Dirac equation itself (Cf.ref.\cite{uof} and \cite{bgs1,bgs2,bgs3}). There is still a convergence of the Classical theory of the electron and its Quantum Mechanical counterpart.\\
Finally we make the following remark: From the above it appears that causality and Special Relativity break down within the Compton scale - Special Relativity as pointed out elsewhere \cite{cu} is valid outside the fuzzy Compton scale, within which space time is ill defined. It is interesting in this connection to note that if within the Compton scale we were to work with the non relativistic theory, then as is well known \cite{rohr} we recover the electron radius as,
$$R = \frac{e^2}{mc^2}$$
In other words we get the intertial mass entirely as the electromagnetic mass. Outside the Compton scale, without contradiction Relativity and Relativistic Electrodynamics work, this being the new input.

\end{document}